\documentstyle[tighten,preprint,eqsecnum,aps,prd]{revtex}
\def\no{\nonumber}
\def\a{\alpha}
\def\be{\begin{equation}}
\def\ee{\end{equation}}
\def\p{\partial}
\def\T{\Theta}
\def\r{r}
\def\k{\kappa}
\def\a{\alpha}
\def\t{\theta}
\def\S{\Sigma}

\begin{document}
\draft

\preprint{\vbox{\baselineskip=12pt
\rightline{IUCAA-16/99}
\rightline{}
\rightline{hep-th/9911070}
}}
\title{Quasilocal energy for rotating charged black hole solutions in general
relativity and string theory}
\author{Sukanta Bose\footnote{Electronic address:
{\em sbose@iucaa.ernet.in}} and Thant Zin Naing\footnote{\em Physics 
Department, Yangon University, Myanmar}}
\address{Inter-University Centre for Astronomy and Astrophysics, Post Bag 4,
Ganeshkhind,\\ Pune 411007, India}

\date{April 1999}

\maketitle
\begin{abstract}

We explore the (non)-universality of Martinez's conjecture, originally proposed
for Kerr black holes, within and beyond general relativity. The conjecture 
states that the Brown-York quasilocal energy at the outer horizon of such a 
black hole reduces to twice its irreducible mass, or equivalently, to $\sqrt{A}
/2\sqrt{\pi}$, where $A$ is its area. We first consider the charged Kerr black
hole. For such a spacetime, we calculate the quasilocal energy within a 
two-surface of constant Boyer-Lindquist radius embedded in a constant 
stationary-time slice. Keeping with Martinez's conjecture, at the outer horizon
this energy equals to $\sqrt{A}/2\sqrt{\pi}$. The energy is positive and 
monotonically decreases to the ADM mass as the boundary-surface radius 
diverges. Next we perform an 
analogous calculation for the quasilocal energy for the Kerr-Sen spacetime, 
which corresponds to four-dimensional rotating charged black hole solutions in 
heterotic string theory. The behavior of this energy as a function of the 
boundary-surface radius is similar to the charged Kerr case. However, we show 
that it does not approach the expression conjectured by Martinez at the 
horizon. 

\end{abstract}

\narrowtext

\vfil
\pagebreak

\section{Introduction}

A geometric theory of gravity, such as general relativity, is known to lack a
meaningful notion of local energy density \cite{MTW,ABS,BY}. This is 
essentially due to the absence of an unambiguous prescription for decomposing 
the spacetime metric into ``background'' and ``dynamical'' components. The 
low-energy effective field theory for heterotic string theory is no exception.
Such theories, however, admit several alternative prescriptions for computing
quasilocal energies (see Refs. \cite{BY,BL} and the references therein). In 
Ref. \cite{BY} (henceforth referred to as BY), Brown and York introduced 
a way to define the quasilocal energy (QLE) of a spatially bounded system in 
general relativity in terms of the total mean curvature of the boundary. In 
this paper, we first apply this definition to the case of charged Kerr family 
of black hole spacetimes, which represent rotating, charged black hole 
solutions in general relativity. Next we apply it to study the 
behavior of quasilocal energy of the Kerr-Sen family, which represents 
rotating, charged 
black hole solutions in heterotic string theory \cite{sen1}. Since the static, 
charged (or neutral) solution is a special case of this family, the QLE of such
a solution is obtained in the limit of vanishing angular momentum (or charge).
We perform these calculations in the regime of the slow-rotation approximation.

The motivation for the analysis in this paper is as follows. It is of interest 
to explore the form of the classical laws of black hole mechanics and the 
ensuing picture of black hole thermodynamics in alternative theories of 
gravity such that they can be compared with the corresponding scenario in
general relativity. But the study of the thermodynamical laws entails the
knowledge of the energy and entropy associated with these spacetimes. Moreover,
equilibrium thermodynamics of a black hole (specifically, in the case of 
an asymptotically flat solution), requires that it be put in a finite-sized 
``box'', just as one does in general relativity \cite{BPP1}. Thus, such a study
requires the knowledge of the quasilocal energy of these ``finite-sized'' 
systems. Moreover, in a study of the quasilocal energy corresponding to 
different types of two-boundaries embedded in constant stationary-time slice
of the Kerr spacetime (in the slow-rotation approximation), Martinez has 
conjectured that the QLE approaches twice the irreducible mass of such a black
hole. In this paper, we study the (non)-universality of this conjecture. We 
find that it remains valid for the charged Kerr black hole. In the case of the 
Kerr-Sen black hole, however, we show that the QLE evaluated on a two-surface 
of constant Boyer-Lindquist radius embedded in a constant stationary-time slice
does not approach the expression conjectured by Martinez at the horizon. 

We introduce the BY quasilocal energy in section \ref{sec:QE}. We use this 
definition in section \ref{sec:KN} to find the QLE of the charged Kerr black 
hole. Similarly, in section \ref{sec:KS} we evaluate the QLE of the Kerr-Sen
black hole. We present our observations on the results of these sections and 
discuss the status of Martinez's conjecture for these two cases 
in section \ref{sec:disc}. Finally, in appendix \ref{app} we derive the
integral expression for the QLE associated with a certain class of quasilocal
two-surfaces embedded in an axisymmetric spatial three-slice. This expression
is extensively used in this paper for the QLE computations for different 
spacetimes mentioned above. We follow the conventions of Ref. \cite{MTW} and 
employ geometrized units $G=1=c$.

\section{Brown-York Quasilocal Energy}
\label{sec:QE}


The BY derivation of the quasilocal energy, as applied to a four-dimensional 
(4D) spacetime solution of Einstein gravity can be summarized as follows. The 
system one considers is a 3D spatial hypersurface $\Sigma$ bounded by a 2D
spatial surface ${\sf B}$ in a spacetime region that can be decomposed as a 
product of a 3D hypersurface and a real line-interval representing time. 
The time-evolution of the boundary ${\sf B}$ is the surface ${}^3 {\sf B}$. 
One can then obtain a
surface stress-tensor on ${}^3 {\sf B}$ by taking the functional derivative of
the action with respect to the 3D metric on ${}^3 {\sf B}$. The energy surface 
density is the projection of the surface stress-tensor normal to a family of 
spacelike surfaces such as ${\sf B}$ that foliate ${}^3 {\sf B}$.
The integral of the energy surface density over such a boundary ${\sf B}$ is
the quasilocal energy associated with a spacelike hypersurface  $\Sigma$ whose
{\em orthogonal} ~intersection with ${}^3 {\sf B}$ is the boundary ${\sf B}$.
Here we assume that there are no inner boundaries, such that the spatial
hypersurfaces $\Sigma$ are complete. In the case where horizons form, one
simply evolves the spacetime inside as well as outside the horizon.
Under these conditions, the QLE is defined as:
\be \label{BYE}
 E = \frac{1}{8 \pi} \oint_{\sf B} d^2 x \sqrt{\sigma} (k - k_0) \ \ ,
\ee
where $\sigma$ is the determinant of the 2-metric on ${\sf B}$, $k$ is the
trace of the extrinsic curvature of ${\sf B}$, and $k_0$ is a reference term
that is used to normalize the energy with respect to a reference spacetime, not
necessarily flat. To compute
the QLE for asymptotically flat solutions, we will choose the reference
spacetime to be flat as well. In that case, $k_0$ is the trace of the extrinsic
curvature of a two-dimensional surface embedded in flat spacetime, such that it
is isometric to ${\sf B}$. 

Interestingly, the foregoing analysis can be applied in a straightforward way
to compute QLE of spatial sections of solutions of scalar-tensor theories as
well. This is because, for spacetime dimensions higher than two, a 
scalar-tensor theory can be cast in the ``Einstein-Hilbert'' form by performing
a conformal transformation. The solutions of this conformally related action 
are given by the Einstein-frame metrics, which themselves are related to 
solutions of the scalar-tensor theory by a conformal transformation. It was 
shown in Ref. \cite{BL} that the quasilocal mass (QLM), which is a construct 
related to QLE, is invariant under such a transformation. When there is a 
timelike Killing vector field $\xi^\mu$ on the boundary ${}^3 {\sf B}$, such 
that it is also hypersurface forming, one can define an associated conserved 
QLM for the bounded system 
\cite{BCM,BY}:
\begin{equation}
\label{massaux}
M = \int_{\sf B} d^{(D-1)} x \> \sqrt{\sigma} \> N \varepsilon 
\ \ ,
\end{equation}
where $N$ is the lapse function related to $\xi^\mu$ by $\xi^\mu = Nu^\mu$. 
Further, if $\xi \cdot u = -1$, then $N=1$ and 
consequently the QLM is the same as the QLE. Unlike QLE, 
the quasilocal mass is independent of any foliation of the bounded system.
Moreover, owing to its conformal invariance, a frame of convenience can be 
chosen for the computation of QLM without affecting its value. The QLE, 
in general, is not a conformal invariant. Thus, the 
frame in which it is evaluated needs to be clearly specified. In this paper, 
all computations of the QLE will be carried out in the Einstein frame. This is 
the frame in which Sen analyzed the properties of his rotating charged black 
hole solution. 

\section{QLE for charged Kerr black holes}
\label{sec:KN}

The charged Kerr solution in general relativity represents the spacetime of a 
rotating, charged black hole. Its spacetime metric and electromagnetic vector
potential are given by \cite{KN}
\begin{eqnarray}
ds^2=&-&\left(\frac{\Delta -a^2 \sin^2}{\S}\right) dt^2- \frac{2a\sin^2\theta
(r^2 +a^2 -\Delta)}{\S} dt d\phi \nonumber \\ 
&+&\left[\frac{(r^2+a^2)^2-\Delta a^2\sin^2\theta}{\S}\right]\sin^2\theta 
d\phi^2+\frac{\S}{\Delta} d\r^2+\S d\theta^2  \ \ , \label{knmet}\\
A_a = &-&{Qr\over\S}[ (dt)_a - a \sin^2\theta (d\phi)_a ] \ \ ,
\end{eqnarray}
where
\begin{eqnarray}
\S &=& r^2+a^2 \cos^2 \theta \ \ , \\
\Delta &=& r^2+a^2 +Q^2-2 M r \ \ , 
\end{eqnarray}
The above fields define a black hole solution with mass $M$, charge $Q$,
and specific angular momentum $a$.

\subsection{Unreferenced QLE for charged Kerr black holes}

Consider a constant stationary-time hypersurface $\S$ embedded in the 
charged Kerr spacetime with the metric (\ref{knmet}). Define the two-surface 
${\sf B}$ to be a surface with some constant value for the 
Boyer-Lindquist radial coordinate, $r=r_0$, embedded in $\S$. The line-element 
on this surface in the slow-rotation approximation, i.e., $a\ll r_0$, is:
\be
ds^2 \approx  r_0^2\Bigl( 1 + {{a^2}\over{r_0^2}} \cos^2 \theta \Bigr)d\theta^2
+r_0^2\Biggl\{1 + {{a^2}\over{r_0^2}}\left[1 + \left({{2M}\over{r_0}}-{Q^2
\over r_0^2}\right)\sin^2\theta\right]\Biggr\}\sin^2\theta\, d\phi ^2 \,. 
\label{knBmets}
\ee
Terms of order $O(a^4/\r_0^4)$ and higher have been neglected.

We now calculate the unreferenced QLE at the two-surface ${\sf B}$, 
defined by $r= r_0$. We assume that $r_0 \geq r_+$, where $r_+$ represents the 
outer horizon of the charged Kerr black hole. Using Eq. (\ref{int}) of the 
appendix, the unreferenced QLE $\varepsilon$ for the surface $r=r_0$ can be 
written explicitly as
\begin{eqnarray}
\varepsilon  &=&
-{{{r_0}} \over 2} \sqrt{ 1 - {{2M}\over{{r_0}}} + {{a^2+Q^2}\over{{r_0} ^2}} }
\times\nonumber \\
& &\int_{0} ^{\pi} d\theta \sin \theta \Bigg\{ 1 - {{a^2}\over{2{r_0} ^2}} 
\bigg[\cos ^2 \theta +\bigg( {{M}\over{{r_0}}}- {{Q^2}\over{{r_0^2}}}\bigg)
\sin^2\theta \bigg] + O \bigg({{a^4}\over{{r_0} ^4}}\bigg)\Bigg\} \ ,
\end{eqnarray}
where we have retained only terms of leading order in the parameter $a/r_0$. 
The above integration can be performed in a straightforward way to yield:
\begin{equation}\label{knuqle}
\varepsilon = -{r_0} \sqrt{ 1 - {{2M}\over{{r_0}}}+{{a^2+Q^2}\over{{r_0} ^2}} }
\, \, \Bigg[1 - {{a^2}\over {6{r_0}^2}}\bigg(1 +{{2M}\over{{r_0}}} -{{Q^2}\over
{{r_0^2}}}\bigg)+O \bigg({{a^4}\over{{r_0} ^4}} \bigg)\Bigg]\ \, \label{knue}
\end{equation}
Note that in the limit $Q \to 0$, the above expression simplifies to
\be\label{Q0E}
\varepsilon=-r_0 \sqrt{1-\frac{2M}{r_0}+\frac{a^2} {r_0^2}}\left[1- \frac{a^2}
{6r_0^2} \left(1+\frac{2 M}{r_0}\right) +O\left(\frac{a^4}{r_0^4}\right)
\right] \ \ ,
\ee
which is the unreferenced QLE of the neutral Kerr black hole \cite{EM}. It also
has the expected limit when $a \to 0$, in which case Eq. (\ref{knuqle}) gives
\be\label{ka0E}
\varepsilon = -{r_0} \sqrt{ 1 - {{2M}\over{{r_0}}}+{{Q^2}\over{{r_0} ^2}} }
\ \ ,
\ee
which is the unreferenced QLE of the Reissner-Nordstrom black holes \cite{BY}.

Note that no approximations have been made inside the square-root appearing in 
$\varepsilon$. As $r_0\to \infty$,  we have $\varepsilon \to M-\r_0$,
which is divergent. This prompts the need for a subtraction term to 
renormalize the unreferenced QLE. Below, we compute such a reference term.

\subsection{Reference term}
\label{subsec:knref}

To obtain the reference term in the QLE expression, Eq. (\ref{BYE}), we first 
find a 2D surface isometric to (\ref{knBmets}), which is 
embeddable in a flat 3D slice with the line element
\begin{equation}
ds^2 = d{{\cal R}}^2 + {{\cal R}}^2 d{\Theta}^2 +{{\cal R}}^2
\sin ^2 \Theta \, d\Phi^2 \ \ ,
\label{flat}
\end{equation}
where ${\cal R}$, ${\Theta}$, and $\Phi$ are the spherical polar coordinates.
Let the desired 2D surface in the flat slice be characterized by ${\cal R} = 
f(\Theta)$, where $f$ is a function of the azimuthal angle $\Theta$ and the 
parameters $(M,a,{r_0})$. Its intrinsic metric is  obtained from (\ref{flat})
as follows. We assume that $\Theta = \T (\t)$ and $\Phi = \phi$. Then, on the 
two-surface ${\sf B}$, we find that ${\cal R}$ is a function of $\t$, i.e., 
${\cal R}= R(\t)$, say. Hence, the line-element on ${\sf B}$ is 
\be\label{knBmet1}
ds^2 = [\dot{R}^2 + R^2\dot{\T}^2]d\t^2 + R^2\sin^2\T d\phi^2 \ \ ,
\ee
where an overdot denotes derivative with respect to $\t$.

Requiring the above line-element to be isometric to (\ref{knBmets}) implies 
that the following couple of equations be obeyed:
\be\label{kntt}
\dot{R}^2+R^2\dot{\Theta}^2=r_0^2\left[1+\frac{a^2}{r_0^2}\cos^2\Theta\right] 
\ \ ,\ee
and
\be\label{knpp}
R^2 \sin^2 \Theta=r_0^2 \sin^2 \theta\left[1+\frac{a^2}{r_0^2}\left(1+\frac{2M}
{r_0}\sin^2 \theta-\frac{Q^2}{r_0^2}\sin^2\theta \right)\right] \,. 
\ee
Assuming that $\dot{R}^2 \simeq O(a^4/r_0^4 )$ (this condition will be 
justified {\em a posteriori}), the above equations can be combined to yield 
the following first-order ordinary differential equation:
\be\label{knode}
\frac{d{\Theta}}{\sin\Theta}=\frac{d\theta}{\sin\theta}\left[1-\frac{a^2}
{2r_0^2}\sin^2\theta \left(1+\frac{2M}{r_0}-\frac{Q^2}{r_0^2}\right)\right] 
\ \ ,\ee
which is easily solved to give:
\be\label{knsint}
\sin\T = \sin\t 
\left[1+ {a^2\over 2r_0^2}\left(1+{2M\over r_0} - {Q^2\over r_0^2}\right)
\cos^2\t\right]
\ee
Putting this back in Eq. (\ref{knpp}), we find $R(\t)$. In the resulting 
expression, using Eq. (\ref{knsint}) to substitute for $\t$ in terms of $\T$, 
we finally get 
\be\label{knfT}
f(\T)=r_0\left[ 1+{a^2\over 2r_0^2} \sin^2\T -{a^2\over 2r_0^2}\left({2M\over 
r_0}-{Q^2\over r_0^2}\right) \cos^2\T\right] \,.
\ee
The two-surface ${\cal R}= f(\T)$ then describes an oblate spheroid, which 
bulges out near the equator (i.e., near $\T = \pi/2$). Note that the above 
equations can be used to prove that indeed $\dot{R}^2 \simeq O(a^4/ \r_0^4)$.

\subsection{Referenced QLE for charged Kerr black hole}

In the slow-rotation approximation the intrinsic metric on ${\sf B}$,
as embedded in flat space, is
\begin{equation}
ds^2 \approx r_0^2 \left[ 1+{a^2\over r_0^2} \sin^2\T -{a^2\over r_0^2}\left(
{2M\over r_0}-{Q^2\over r_0^2}\right)\cos^2\T\right]\>\left( d{\Theta}^2+ 
\sin^2 \Theta \, d\Phi^2 \right) \ . \label{knBmet2}
\end{equation}
The extrinsic curvature $k_0$ of this surface embedded in flat space can be 
evaluated using the method detailed in the appendix. This in turn can be used 
to compute the renormalization integral in the QLE (\ref{BYE}):
\be
\varepsilon^0 = -r_0\left[1+\frac{a^2}{3 r_0^2}\left(1+\frac{M}{r_0}-\frac{Q^2}
{2r_0^2}\right)\right]\,.
\ee
The referenced QLE is, therefore, obtained to be:
\begin{eqnarray}
E = &&{r_0}\left[1 -\sqrt{1-\frac{2M}{r_0}+\frac{a^2+Q^2} {r_0^2}}
\right]+ {a^2\over 6r_0} \Bigg[ 2\left( 1 + {M\over r_0}-{Q^2\over 2r_0^2} 
\right) \no\\
&+& \left(1+\frac{2M}{r_0}-\frac{Q^2} {r_0^2}\right)\sqrt{1-\frac{2M}{r_0}+
\frac{a^2+Q^2} {r_0^2}}\Bigg] \,. \label{knfineE}
\end{eqnarray}
As $r_0\to \infty$, we have 
\be
E\to {r_0}\left\{1 -\left[1-\left({M\over r_0}-{a^2+Q^2\over 2r_0^2}\right)
\right] \right\}\to M + O\left({1\over r_0}\right)\ \ ,
\ee
which is indeed the Arnowitt-Deser-Misner (ADM) mass \cite{ADM} 
of the charged Kerr spacetime.

Near the outer horizon $r_0 = r_+$, the energy is 
\be
E(r_0 = r_+)= r_+ \left[ 1+{a^2 \over 2r_+^2} \right] \,.
\ee
In this limit,
\be\label{knEr+A}
E(r_0 = r_+) \simeq \left[{1\over 4\pi} A\right]^{1/2} = 2M_{\rm irr}
\left[1+O\left(\frac{a^4}{r_+^4}\right)\right]
\ee
to leading order in $a/r_+$. Above, we have defined 
\begin{equation}
M_{\rm irr}^2=\frac{1}{2}\left[M r_+ - \frac{Q^2}{2}\right]
\end{equation}
for the charged Kerr black hole, which is the generalization of the 
irreducible mass of a neutral Kerr black hole \cite{CR}.

\section{The Kerr-Sen solution}
\label{sec:KS}

Consider the following string-inspired low-energy effective action in four 
dimensions:
\be\label{action}
S=-{1\over 16\pi}\int d^4 x\sqrt{- \det~G} e^{-\Phi} (-R+{1\over
12}H_{\mu\nu\r}H^{\mu\nu\r} -G^{\mu\nu}\p_\mu\Phi\p_\nu\Phi
+{1\over 8}F_{\mu\nu}F^{\mu\nu}) \,.
\ee
Here $G_{\mu\nu}$ is the metric, $R$ is the scalar curvature,
$F_{\mu\nu}=\p_\mu A_\nu-\p_\nu A_\mu$ is the field strength
corresponding to the Maxwell field $A_\mu$, $\Phi$ is the dilaton field, and
\be\label{Hmnr}
H_{\mu\nu\r} =\p_\mu B_{\nu\r} +{~\rm cyclic~permutations}
-(\Omega_3(A))_{\mu\nu\r} \ \ ,
\ee
where $B_{\mu\nu}$ is the antisymmetric tensor gauge field. Above we have 
defined
\be
(\Omega_3(A))_{\mu\nu\r}={1\over 4} (A_\mu F_{\nu\r}+ {~\rm cyclic
{}~permutations}) \ \ ,
\ee
which is the gauge Chern-Simons term. It must be noted that the above theory 
is one where 6 of the 10 dimensions have been compactified to a six-torus.
The massless fields arising from compactification have not been included
in the effective action. Only a $U(1)$ component of the full set of non-abelian
gauge fields present in the theory has been included above. Consequently, 
the corresponding solutions carry a $U(1)$ charge only. The metric $G_{\mu\nu}$
used here is the metric that arises naturally in
the $\sigma$-model, and is related to the Einstein metric
$g_{\mu\nu}$ through the relation:
\be\label{CT}
g_{\mu\nu} = e^{-\Phi} G_{\mu\nu} \,.
\ee
Finally, the action was truncated to contain only those terms 
that contain two or a lesser number of derivatives.

Sen showed that the above theory has rotating charged black hole solutions 
given by the following field configuration \cite{sen1}:
\begin{mathletters}%
\label{sol}
\begin{eqnarray}
ds^2&=& -{(\r^2 +a^2\cos^2\theta -2m\r)
(\r^2+a^2\cos^2\theta) \over (\r^2+a^2\cos^2\theta
+2m\r\sinh^2{\alpha\over 2})^2} dt^2 - {4m\r a\cosh^2{\alpha\over 2} 
(\r^2+a^2\cos^2\theta)\sin^2\theta
\over (\r^2 +a^2\cos^2\theta +2m\r\sinh^2{\alpha\over 2})^2} dt d\phi\no \\
&& +{\r^2 +a^2\cos^2\theta\over \r^2
+a^2 -2m\r} d\r^2 + (\r^2 +a^2\cos^2\theta) d\theta^2\no\\
&&+\{(\r^2+a^2)(\r^2+a^2\cos^2\theta) +2m\r a^2\sin^2\theta +4m\r
(\r^2+a^2) \sinh^2{\alpha\over 2}+ 4m^2\r^2\sinh^4{\alpha\over 2}\} \no\\
&& \qquad \times {(\r^2+a^2\cos^2\theta)\sin^2\theta\over (\r^2
+a^2\cos^2\theta +2m\r\sinh^2{\alpha\over 2})^2} d\phi^2 \ \ , \label{ksmet} \\
\Phi &=& -\ln {\r^2 +a^2\cos^2\theta +2m\r\sinh^2{\alpha\over 2} \over
\r^2 +a^2\cos^2\theta} \ \ , \\
A_\phi &=& -{2m\r a\sinh\alpha\sin^2\theta\over \r^2 +a^2\cos^2\theta
+2m\r\sinh^2{\alpha\over 2}} \ \ , \\
A_t &=& {2m\r\sinh\alpha\over \r^2 +a^2\cos^2\theta
+2m\r\sinh^2{\alpha\over 2}} \ \ , \\
B_{t\phi} &=& {2m\r a\sinh^2{\alpha\over 2}\sin^2\theta \over \r^2
+a^2\cos^2\theta +2m\r\sinh^2{\alpha\over 2}} \,.
\end{eqnarray}
\end{mathletters}%
The other components of $A_\mu$ and $B_{\mu\nu}$ vanish.
The Einstein metric $ds^2_E\equiv e^{-\Phi}ds^2$ is given by
\begin{eqnarray}
ds_E^2&=&-\left(1-\frac{2 m \r \cosh^2 {\alpha\over 2}}{\Upsilon}\right) dt^2+
\frac{\Upsilon}{\Gamma} d\r^2+\Upsilon d\theta^2 - \frac{4 m \r a\cosh^2
{\alpha\over 2}\sin^2 \theta}{\Upsilon} dt d\phi \nonumber \\ 
&&+\left[\frac{(\r^2+a^2+2 m \r \sinh^2{\alpha\over 2})^2-\Gamma a^2\sin^2
\theta}{\Upsilon}\right]\sin^2\theta d\phi^2 \ \ , \label{Emet}
\end{eqnarray}
where
\begin{eqnarray}
\Upsilon &=& \r^2+a^2 \cos^2 \theta+2 m \r \sinh^2{\alpha\over 2} \ \ ,\no\\
\Gamma &=& \r^2+a^2 -2 m \r  \,. \no
\end{eqnarray}
This metric describes a black hole solution with mass $M$, charge $Q$,
angular momentum $J$, and magnetic dipole moment $\mu$ given by,
\be\label{para}
M={m\over 2} (1+\cosh\alpha),~~~~ Q={m\over\sqrt 2}\sinh\alpha, ~~~~ J=
{ma\over 2} (1+\cosh\alpha),~~~~ \mu ={1\over\sqrt 2} ma\sinh\alpha \,.
\ee
This is often called the Kerr-Sen black hole solution. The location of the
horizon is given by the coordinate singularities, which occur on the surfaces
\be
\r^2 -2m\r +a^2 =0 \,.
\ee
This has the following two roots:
\be\label{horizons}
\r =m\pm \sqrt{m^2-a^2} = M-{Q^2\over 2M}\pm\sqrt{(M-{Q^2\over 2M})^2
-{J^2\over M^2}}\equiv \r_\pm \,.
\ee
which correspond to the outer and inner horizons, respectively.
The area of the outer event horizon with the metric given in Eq. (\ref{Emet})
is found to be
\be
A=8\pi M \left[M-{{Q^2\over 2M}} +\sqrt{\left(M-{Q^2\over 2M}\right)^2
-{J^2\over M^2}} \>\right] \,.
\ee
Equation (\ref{horizons}) shows that the horizon disappears unless
\be
|J|\le M^2-{Q^2\over 2} \,.
\ee
Thus the extremal limit of the black hole corresponds to
$|J|\to M-Q^2/2M$. In this limit, $A\to 8\pi |J|$. (We
note that rotating black hole solutions in a related theory of dilaton gravity
were also found in Ref. \cite{HORO}. )

\subsection{Unreferenced QLE for Kerr-Sen black hole}

Consider a constant stationary-time hypersurface $\S$ embedded in the Kerr-Sen 
spacetime with the metric (\ref{ksmet}). Define the two-surface ${\sf B}$ 
to be the surface with some constant value for the Boyer-Lindquist radial 
coordinate $\r= \r_0$, embedded in $\S$. The line element on this surface in 
the slow-rotation approximation, i.e., $a\ll \r_0$, is:
\be\label{Bmet}
ds^2 = \r_0^2\k \left[1+{a^2\cos^2\t\over \r_0^2\k}\right]d\t^2 + z^2 d\phi^2
\ \ , \ee
where 
\be\label{z}
z= \r_0\sqrt{\k} \sin\t \left\{ 1+{a^2\over 2\r_0^2\k} + {ma^2\over \r_0^3\k^2}
\cosh^2 {\a\over 2} \sin^2\t \right\} \ \ ,
\ee
and $\k=1+(2m/\r_0) \sinh^2\alpha/2$. Terms of order $O(a^4/\r_0^4)$ 
and higher have been neglected.

We now calculate the unreferenced QLE within the two-surface ${\sf B}$, 
defined by $\r= \r_0$.  We assume that $\r_0 \geq \r_+$, where 
$\r_+$ represents the outer horizon of the Kerr-Sen black hole.  Using 
Eq. (\ref{int}) of the appendix, the
integral $\varepsilon$ for the surface $\r=\r_0$ can be written explicitly as
\begin{eqnarray}
\varepsilon=-\frac{\r_0 \k'}{2\k}\sqrt{1-\frac{2 m}{\r_0}+\frac{a^2}{\r_0
^2}}
\int_0^\pi d\theta &&\sin\theta \Bigg[1-\frac{a^2}{2\r_0^2 \k}\left(\cos^2
\theta +\frac{m}{\r_0 \k}\cosh^2 {\alpha \over 2} \right) \nonumber \\
&+&\frac{m^2 a^2}{2\r_0^4 \k^2 \k'}\sinh^2{\alpha \over 
2}\cosh^2{\alpha \over 2}\sin^2\theta +O\left({a^4\over \r_0^4}\right)\Bigg] 
\ \ ,
\end{eqnarray}
where where $\k'=1+(m/\r_0) \sinh^2\alpha/2$.
On performing the above integration we get:
\begin{eqnarray}
\varepsilon=&-&\frac{\r_0 \k'}{\sqrt{\k}} \sqrt{1-\frac{2m}{\r_0}+
\frac{a^2}{\r_0^2}} \nonumber \\
&\times& \left[1- \frac{a^2}{6 \r_0^2 \k} \left(1+\frac{2 m}{\r_0 \k} 
\cosh^2
{\alpha \over 2} -\frac{2 m^2}{\r_0^2 \k \k'}\sinh^2 {\alpha \over 2} \cosh^2
{\alpha \over 2}\right)+O\left(\frac{a^4}{\r_0^4}\right)\right] \ \ , 
\label{unrefE}
\end{eqnarray}
which is the unreferenced QLE for Kerr-Sen black hole.

In the limit $\alpha \to 0$, the expression (\ref{unrefE}) gives
\be\label{alpha0E}
\varepsilon=-\r_0 \sqrt{1-\frac{2m}{\r_0}+\frac{a^2}{\r_0^2}}\left[1-
\frac{a^2}{6 \r_0^2} \left(1+\frac{2 m}{\r_0 } +O\left(\frac{a^4}{\r_0^4}
\right)\right)
\right] \,.
\ee
This is the unreferenced QLE of the neutral Kerr black hole \cite{EM}. It also
has the expected limit when $a \to 0$, in which case
\be\label{a0E}
\varepsilon=-\frac{\r_0 \k'}{\sqrt{\k}}\sqrt{1-\frac{2m}{\r_0}+\frac{a^2} 
{\r_0^2}} \ \ ,
\ee
which is the unreferenced QLE of the static, charged dilatonic black holes
\cite{BL}.

As $\r_0\to \infty$,  we have $\varepsilon \to M-\r_0$, which is again
divergent. Below, we compute the reference term required to renormalize 
this QLE. 

\subsection{Two-surface isometric to ${\sf B}$ embedded in flat space}

To obtain the reference term in the QLE expression, Eq. (\ref{BYE}), we first 
find a 2D surface isometric to (\ref{Bmet}), which is 
embeddable in a flat 3D slice with the line element
\begin{equation}
ds^2 = d{{\cal R}}^2 + {{\cal R}}^2 d{\Theta}^2 +{{\cal R}}^2
\sin ^2 \Theta \, d\Phi^2 \ .
\label{3flat}
\end{equation}
The equation for the desired 2D surface in the flat slice is
denoted  by ${\cal R} = g(\Theta)$, where $g$ is a function of the azimuthal
angle $\Theta$ and the parameters $(M,a,{\r_0})$ of the surface in $\Sigma$. 
Its intrinsic metric is  obtained from (\ref{3flat}). We assume that $\Theta =
\T (\t)$ and $\Phi = \phi$. Then, on the two-surface ${\cal R}$ is a function 
of $\t$, i.e., ${\cal R}= R(\t)$, say. Hence, the line-element on it is given 
by Eq. (\ref{knBmet1}).

Similar to the case of the charged Kerr black hole discussed in section 
\ref{subsec:knref} , even for the Kerr-Sen spacetime, requiring the 
line-element (\ref{knBmet1}) to be isometric to (\ref{Bmet}) yields the 
following first-order ordinary differential equation:
\be\label{ode}
{\dot{\T} \over \sin \T} = {\r_0\sqrt{\k}\over z} \left[1+{a^2cos^2\t\over 2
\r_0^2 \k}\right] \,.
\ee
This equation is easily solved to give:
\be\label{sint}
\sin\T = \sin\t 
\left[1+ {a^2\over 2\k\r_0^2}\left(1+{2m\over \k\r_0} \cosh^2 {\a\over 2}
\right) \cos^2\t\right] \,.
\ee
As in the case of charged Kerr, the above expression can be used to find 
\be\label{gT}
g(\T) = \sqrt{\k}\r_0 \left[ 1+{a^2\over 2\k\r_0^2} \sin^2\T -{ma^2\over k^2
\r_0^3}\cosh^2{\a\over 2} \cos^2\T\right] \,.
\ee
The two-surface ${\cal R}= g(\T)$ once again describes an oblate spheroid.


\subsection{The referenced QLE}

In the slow-rotation approximation, Eq. (\ref{gT}) implies that the intrinsic 
metric on ${\sf B}$, as embedded in flat space, is
\begin{equation}
ds^2 \approx \sqrt{\k}\r_0 \left[ 1+{a^2\over 2\k\r_0^2} \sin^2\T -{ma^2\over 
\k^2\r_0^3} \cosh^2{\a\over 2} \cos^2\T\right]\>\left( d{\Theta}^2 + \sin ^2 
\Theta \, d\Phi^2 \right) \ . \label{Bmet2}
\end{equation}
The extrinsic curvature $k_0$ of this surface embedded in flat space can be 
evaluated using the method detailed in the appendix. This in turn can be used 
to compute the renormalization integral in the QLE (\ref{BYE}):
\begin{eqnarray}
\varepsilon^0 &=&{1\over\k}\int_{{\sf B}}k^0 \sqrt{\sigma} d\T d\Phi\no\\
&=& -\r_0 \sqrt{\k}\left[ 1 + {a^2\over 3\k\r_0^2} \left( 1 + {m\cosh^2 {\a
\over 2} \over \k\r_0} \right)\right] \,.\label{bye0}
\end{eqnarray}
The referenced QLE is, therefore, obtained to be:
\begin{eqnarray}
E = &&{\r_0\over\k}\left[\k -\k'\sqrt{1-\frac{2m}{\r_0}+\frac{a^2} {\r_0^2}
}\right] + {a^2\over 6\r_0\k^{3/2}}\Bigg[ 2\k\left( 1 + {m\cosh^2 {\a\over 2} 
\over \k\r_0} \right) \no\\
&+& \k'\sqrt{1-\frac{2m}{\r_0}+\frac{a^2} {\r_0^2}}
\left(1+\frac{2 m}{\r_0 \k} \cosh^2 {\alpha \over 2} -\frac{2 m^2}{\r_0^2 
\k \k'}\sinh^2 {\alpha \over 2} \cosh^2 {\alpha \over 2}\right)+O\left(\frac{
a^4}{\r_0^4}\right)\Bigg] \,. \label{finE}
\end{eqnarray}
As $\r_0\to \infty$, we have 
\be
E\to {\r_0\over\k}\left[\k -\k'\left(1-{m\over \r_0}\right)\right] 
\to m\cosh^2 {\alpha\over 2} \ \ ,
\ee
which is indeed the ADM mass (\ref{para}) of the Kerr-Sen solution.

The QLE (\ref{finE}) has the correct limit for vanishing charge, namely, for
$\alpha =0$. In that case, one obtains the expression in Eq. (\ref{knfineE}),
which is the QLE for the Kerr black hole. Similarly, for vanishing rotation,
i.e., for $a =0$, Eq. (\ref{finE}) reduces to
\be
E = {\r_0\over\k}\left[\k -\k'\sqrt{1-\frac{2m}{\r_0}}\right] \ \ ,
\ee
which is the QLE for charged black holes in string theory \cite{BY}.

Near the outer horizon $\r_0 = \r_+$, the energy is 
\begin{eqnarray}
E(\r_0 = \r_+)&=& \r_+ \left[ 1+{a^2 \over 3\kappa \r_+^2} \left( 1+{\cosh^2 
{\alpha\over 2}\over 2\kappa}\right)+O\left(\frac{a^4}
{\r_+^4}\right)\right] \ \ , \label{Er+} \\
&=& 2m\left\{1-{a^2\over 4m^2}\left[1-{1\over 3\kappa}\left(1+{\cosh^2 
{\alpha\over 2}\over 2\kappa}\right)\right]+O\left(\frac{a^4}
{\r_+^4}\right)\right\} \,. \label{Er+m}
\end{eqnarray}
In the limit of vanishing charge, which is given by $\alpha=0$, the above 
quantity goes over to
\begin{eqnarray}
E(\r_0 = \r_+)&=& \r_+ \left[ 1+{a^2 \over 2 \r_+^2}+O\left(\frac{a^4}
{\r_0^4}\right)\right] \ \ , \label{Er+0} \\
&=& 2m\left[1-{a^2\over 8m^2}+O\left(\frac{a^4}
{\r_0^4}\right)\right] \,. \label{Er+m0}
\end{eqnarray}
which are the expected values for the Kerr black hole \cite{EM}. Note that in 
this latter case,
\be\label{Er+A}
E(\r_0 = \r_+) \simeq \left[{1\over 4\pi} A\right]^{1/2} 
\ee
to leading order in $a/\r_+$. Whether such a relation continues to hold even
after the slow-rotation approximation is dropped, is not known. For the 
Kerr-Sen black hole, however,
\be\label{Ar+}
\left[{1\over 4\pi} A\right]^{1/2} = \r_+\cosh{\alpha\over 2}\left[ 1+{a^2 
\over 2 \r_+^2}+O\left(\frac{a^4} {\r_+^4}\right)\right] \ \ ,
\ee
which is not the same as $E(\r_0 = \r_+)$ given in Eq. (\ref{Er+}) for such a 
black hole. 

\section{Discussion}
\label{sec:disc}

The Brown-York quasilocal energy of a Kerr black hole, for the type of slice
$\S$ and quasilocal surface ${\sf B}$ considered in this paper, has not
been evaluated yet for the exact case (i.e., beyond the slow-rotation 
approximation). One of the main hurdles in this computation is the 
determination of the two-surface ${\cal R} = f(\Theta)$, isometric to 
${\sf B}$, to be embedded in a flat three-slice. It is nevertheless
interesting to explore the slow-rotation regime of such black holes in general
relativity and alternative theories of gravity for the results obtained can 
often tell us about the behavior of certain physical quantities in the exact 
case. One such quantity is the value of QLE at the (outer) horizon of such 
black holes, which for the Kerr black hole approaches twice its irreducible 
mass. In this paper, we find that this result continues to hold even for the
charged Kerr black hole. Thus, within general relativity, this identity appears
to bear a universal quality as far as its applicability to stationary black 
hole solutions is concerned. 

It is known that the mechanical laws of stationary black hole solutions in 
general relativity are extendible also to other alternative theories of 
gravity, especially ones connected with a diffeomorphism invariant Lagrangian. 
One such leading alternative to general relativity is a string-inspired
four-dimensional low-energy effective theory. It is of interest to ask if other
identities related to black hole mechanics in general relativity continue to 
hold for black holes in string theory. This motivated us to
study the status of Martinez's conjecture in the context of the Kerr-Sen family
of black holes, which arise as solutions in heterotic string theory. We first
find the QLE of such black holes for a choice of three-slice $\S$ quasilocal 
surface ${\sf B}$ identical to the ones used to evaluate the QLE of charged
Kerr black holes. This expression is found to have the correct limits when 
the radial coordinate is made to diverge, or when the charge is made to vanish.
However, at the outer horizon, the QLE for such black holes does not reduce to 
twice its irreducible mass. It is important to note that the value of the QLE 
is influenced by the choice of a reference term. In fact, at the outer horizon
it is solely this reference term that contributes to the QLE. It may be, 
therefore, be possible to motivate an alternative reference term that is
concommitant with the applicability of Martinez's conjecture even for such 
black holes. We hope to return to this issue elsewhere.

\acknowledgments
  
We thank Naresh Dadhich and Mohammad Nouri-Zonoz for helpful discussions. One 
of us (TZN) would like to
express his gratitude to IUCAA for the hospitality and financial support, which
he received during his stay in IUCAA.

\appendix\section{QLE on a two-sphere embedded in $\Sigma$}
\label{app}

Here we show how to compute the trace the extrinsic curvature, $k$, 
corresponding to a closed two-surface ${\sf B}$ embedded in a 
3D axisymmetric Riemannian manifold. The result can then be used in Eq.
(\ref{BYE}) to evaluate the energy integrals appearing in several places in 
this paper. An analogous calculation is 
given in Ref. \cite{EM}. It is discussed here for the sake of completeness 
only.

Consider a 3D axisymmetric spacelike hypersurface $\Sigma$ described by the 
line element
\be\label{3met}
h_{ij} dx^i dx^j = x^2 dr^2 + y^2 d \vartheta ^2 + z^2 \, d \varphi ^2 \ ,
\ee
where $x^i=(r, \vartheta, \varphi)$ denote arbitrary coordinates adapted 
to the symmetry. The metric coefficients $x$, $y$, and $z$ depend only on the 
``radial" and ``azimuthal" coordinates $r$ and $\vartheta$, respectively.
An arbitrary 2D axisymmetric surface ${\sf B}$ having the topology of a 
two-sphere, and embedded  in 3D space $\Sigma$ is
defined by the relation  $r = R(\vartheta)$, where $R$ is a function of the
azimuthal angle and the parameters of the solution. Its two-dimensional line
element is 
\be\label{2met}
\sigma _{ab} dx^a dx^b
= (x^2 {R'}^2 + y^2)  d \vartheta ^2 + z^2 \, d \varphi ^2 \ , 
\ee
where a prime  denotes differentiation with respect to the coordinate
$\vartheta$. The functions $x$, $y$, and $z$ in (\ref{2met}) are evaluated
at the two-surface $r = R(\vartheta)$.

Let  $n^{i}$ denote the unit outward-pointing spacelike normal to ${\sf B}$
as embedded in $\Sigma$. Its components are
\be\label{normal}
 (n^r, n^{\vartheta}, n^{\varphi}) = {1 \over {\sqrt{y^2 + x^2 R'^2}}} \big(
y/x\, ,  -x\, R'/y\, ,0 \big) \,. 
\ee
The extrinsic curvature of the two-surface ${\sf B}$ as embedded in 
$\Sigma$ is denoted by $ k_{{\mu\nu}}$. Its trace $k$ can be written as
\be\label{k}
k = -{\partial_\mu (n^\mu \sqrt{h})\over \sqrt{h}} \ \ ,
\ee
where $h$ denotes the determinant of the three-metric $h_{\mu\nu}$.
Using the coordinate-components of the unit normal (\ref{normal}) we find that
the trace is
\be\label{kcc}
{k} = -{{1} \over {xyz}} \bigg[ \big( \alpha \,{\gamma}^{-1/2} \big)_{,r} -
\big( \beta \,{\gamma}^{-1/2} \big)_{,\vartheta} \bigg]
\Bigg|_{{\scriptscriptstyle} r=R(\vartheta)} \ , 
\ee
where $\alpha \equiv  y^2 z $, $\beta \equiv  x^2 z \, R' $,
$\gamma \equiv  y^2 + x^2 R'^2 $, and $\delta \equiv  \ln \gamma $. Its proper 
surface integral yields
\be\label{int}
{1 \over 8\pi} \int_{{\sf B}} k \,\sqrt{\sigma}\, d\vartheta d\varphi =
-{1 \over 4} \int_{0}^{\pi} d \vartheta \,{1 \over {xy}} ( \alpha _{,r} -
\beta_{,\vartheta} - {{\alpha}\over 2} \, \delta_{,r} + {{\beta}\over 2}
\delta_{,\vartheta})\bigg| _{{\scriptscriptstyle}  r=R(\vartheta)} \,. 
\end{equation}
This integral is  evaluated at the surface $r = R(\vartheta)$. Finally note 
that both integrals in Eq. (\ref{BYE}) are of the general form (\ref{int}).
It is, however, important to note that each one of these integrals involves 
different values for the functions $x$, $y$, and $z$. Also, the coordinates
on the two-surface, $\vartheta$ and $\varphi$, which appear in these two 
integrals, may not always have identical definitions.

\end{document}